# Quantum Hall Effect in a CVD-Grown Oxide

Oleksandr Zheliuk[1*], Yuliia Kreminska[2], Qundong Fu[3], Davide Pizzirani[1], Andrew A.L.N. Ammerlaan[1], Ying Wang[2], Sardar Hameed[2], Puhua Wan[2], Xiaoli Peng[2], Steffen Wiedmann[1], Zheng Liu[3], Jianting Ye[2*] & Uli Zeitler[1*]

[1] *High Field Magnet Laboratory (HFML-EMFL), Radboud University, Toernooiveld 7, 6525 ED Nijmegen, The Netherlands*

[2] *Device Physics of Complex Materials, Zernike Institute for Advanced Materials, University of Groningen, Nijenborgh 4, 9747 AG, Groningen, The Netherlands*

[3] *School of Materials Science and Engineering, Nanyang Technological University, 50 Nanyang Avenue, Singapore 637371, Singapore.*

**Abstract**

**Two-dimensional electron systems (2DES) are promising for investigating correlated quantum phenomena. In particular, 2D oxides provide a platform that can host various quantum phases such as quantized Hall effect, superconductivity, or magnetism. The realization of such quantum phases in 2D oxides heavily relies on dedicated heterostructure growths. Here we show the integer quantum Hall effect achieved in chemical vapor deposition grown $Bi_2O_2Se$ - a representative member of a more accessible oxide family. A single or few sub-band 2DES can be prepared in thin films of $Bi_2O_2Se$, where the film thickness acts as the sole design parameter and the sub-band occupation is determined by the electric field effect. This new oxide platform exhibits characteristic advantages in structural flexibility due to its layered nature, making it suitable for scalable growth. The unique small mass distinguishes $Bi_2O_2Se$ from other high-mobility oxides, providing a new platform for exploring quantum Hall physics in 2D oxides.**

[*]Corresponding authors. E-mail: oleksandr.zheliuk@ru.nl; j.ye@rug.nl; uli.zeitler@ru.nl

The Quantum Hall effect (QHE), known for its quantized Hall conductance in the units of $\frac{e^2}{h}$, is a fascinating quantum phenomenon that has been actively explored in two-dimensional electron gases (2DEG) and beyond[1,2]. Achieving this macroscopic quantum state typically begins with a meticulous preparation of 2DEG systems pursuing high enough carrier mobility to ensure Landau quantization and sufficiently large potential fluctuations to localize the conduction within the channel width. In traditional semiconductors like Si and GaAs[3], this is often realized by preparing artificial structures like two-dimensional (2D) quantum wells using advanced thin-film technologies, an approach that allows precise control of defects and doping concentrations.

In more recent years, oxides have emerged as an alternative platform capable of realizing both integer and fractional quantum Hall effects[4–7]. Being chemically robust, oxides are promising candidates for constructing stable heterostructures. Physically, 2DEGs based on oxides further introduce interesting possibilities where mobile carriers can be chosen from bands composed mainly of *s*- and *p*-orbitals, such as in ZnO, or from *d*-orbitals as in $SrTiO_3$. Therefore, transport properties benefit from high mobility[8], leading to quantum Hall states, as well as electron correlation effects[7,8]. This facilitates interaction-driven quantum phenomena like magnetism and superconductivity, which can seamlessly interface with the QHE[9,10]. Despite the structural stability and flexibility in hosting various quantum phases in oxide 2DEGs, in comparison to traditional semiconductors, achieving the quantum Hall effect in oxides is constrained by intrinsic factors. Notably, the effective carrier mass of typical oxide 2DEGs is considerably larger, typically falling in the range of 0.3 ~ 1 $m_0$, where $m_0$ is the free electron mass. For example, when searching for quantum Hall effects in the extensively studied interfaces and heterostructures of $SrTiO_3$, the relevant carrier density covers sub-bands coming from the anisotropic Ti, $3d_{xy}$ and $3d_{yz}/3d_{xz}$ orbitals, having masses of ~0.7 and 10 $m_0$, respectively[11]. Consequently, despite achieving ultrahigh quality—where the mobility of thin film $SrTiO_3$[8,7] surpasses that of bulk single crystals—quantizing edge states remains a challenge due to the heavy electrons.

The recently discovered $Bi_2O_2Se$ offers the potential to provide an easy solution to the challenges of quantized transport in oxides. $Bi_2O_2Se$ is a bismuth-based oxychalcogenide, which is synthesized by widely accessible and scalable chemical vapor deposition (CVD)[12–14]. $Bi_2O_2Se$ single crystals are constructed by sharing a Se layer between two Bi-O layers (see Fig. 1a), lacking a distinct Van der Waals (VdW) gap. Although not cleavable as 2D materials, physically, $Bi_2O_2Se$ is a layered semiconductor with an indirect bandgap of ~ 0.8 eV and the transition is from *X* to *Γ* points of the Brillouin zone. The highly dispersive parabolic *Γ* pocket of the conduction band originates mainly from the *p*-orbitals of Bi, leading to a low effective mass of $m^* = 0.14\ m_0$[15]. As carriers mainly navigate within the Bi-O layer in the crystal structure, the electrical transport is effectively encapsulated by the Se layer, essentially forming a buried interface. Consequently, $Bi_2O_2Se$ is insensitive to reactions in ambient conditions and adsorptions as contaminants. This stands in sharp contrast to other prominent systems based on 2D materials, such as black phosphorous[16], where encapsulation as heterostructures is essential to either protect air-unstable materials from reactions or isolate the sensitive surface from contaminations. Moreover, in contrast to conventional semiconductors and oxides, the structural flexibility inherent in layered compounds offers distinct advantages. The layered structure facilitates the incorporation of oxide layers with diverse properties[17], thereby enabling the introduction of additional quantum states, such as superconductivity[18]. Our demonstration of the quantum Hall effect (QHE) in $Bi_2O_2Se$ showcases the potential of this oxide system for readily accessing the quantum Hall state. This opens avenues for further exploration of quantum transport and the development of devices on a resilient oxide platform capable of hosting various quantum phases.

In this paper, we demonstrate that the integer quantum Hall effect can be achieved in CVD-grown $Bi_2O_2Se$ samples of 6 to 30 nm in thickness. In contrast to the layered quasi-2D systems in bulk form, such as transition metal dichalcogenides (TMDs)[19], the Fermi surface of bulk $Bi_2O_2Se$ is a 3D ellipsoid[20]. Although the highest mobility has reached $\sim 10^4 - 10^5$ cm$^2$ V$^{-1}$s$^{-1}$ [20–22] and carrier concentrations as low as $10^{18} - 10^{19}$ cm$^{-3}$ in bulk single crystals, the QHE in 2D $Bi_2O_2Se$ has remained elusive. In our study, we demonstrate that by introducing quantum confinement to form a 2DEG, the QHE is readily observed in samples with one or two orders lower mobility (~1000 cm$^2$ V$^{-1}$s$^{-1}$), making $Bi_2O_2Se$ an excellent candidate for oxide quantum electronics.

**Shubnikov-de Haas oscillation in $Bi_2O_2Se$ field-effect transistors**

The $Bi_2O_2Se$ crystal, as shown in Fig. 1a, has unit cells in a so-called 2D zipper structure[20,23], where the chalcogen plane is shared between two neighboring covalently-bonded bismuth-oxide layers. Isolating Bi-O layers randomly unzip the single Se layer between two newly formed surfaces[20,23]. Once cleaved, the half-covering Se top layer is reconstructed, forming high-density defects. Therefore, we study the as-grown $Bi_2O_2Se$ crystals directly from CVD synthesis. Fig. 1b shows our typical field-effect transistor (FET) device, gated by 300 nm thick $SiO_2$ as the gate dielectric. The thickness of each $Bi_2O_2Se$ channel was extracted from atomic force microscopy (AFM) imaging preceding the device fabrication (Extended Data Fig.1).

All as-grown flakes measured are intrinsically conducting at room temperature. For flakes thicker than ~30 nm, finite bulk conductivity persists down to 4.2 K (Fig. 1c). The subsequently measured transfer characteristics of flakes show distinct thickness dependences. The thick flakes (32 and 60 nm) are intrinsically metallic, where the high conductivities ($\sigma_s \approx 1$ and 0.01 S at 4.2 K, respectively) show no or little response to field-effect gating. When a magnetic field $B$ is applied to thick flakes with high conductivity (*e.g.* the 32 nm sample), the magnetoresistance (MR) varies, with conventional metallic behavior, as $\sim B^2$ (Fig. 1d). The electron concentrations of the different samples determined from the slope of the Hall resistance (Fig. 1e) are in the range of $10^{19}$~$10^{20}$ cm$^{-3}$, consistent with typical values expected for a degenerately doped semiconductor.

On the other hand, the conductivity of thinner samples is significantly reduced (inset of Fig. 1c) and becomes vanishingly small below $V_{th}$, the threshold voltage. The overall Hall response increases with the decrease in thickness (Fig. 1e) indicating lower carrier densities induced in thinner samples. Nevertheless, metallic states, having $\sigma_s$ higher than $\frac{2e^2}{h}$, can be quickly switched on at $V_G > V_{th}$. The gradual decrease of the ON state conductivity upon thinning down (Fig. 1c) is accompanied by a vanishing $\sim B^2$ behavior in the MR (Fig. 1d). Meanwhile, on top of a slowly varying MR background, we observe prominent Shubnikov-de Haas (SdH) oscillations in all measured samples (Fig. 1d and Extended Data Fig. 2). As a finer detail, rich beating patterns, notable in thick samples (*e.g.* 60, 32, and 26 nm), gradually disappears in thinner samples. The variation of oscillation frequency based on thickness can be captured by comparing the fast Fourier transform (FFT) spectra of the SdH signal as a function of $\frac{1}{B}$. Illustrated in Fig. 1f, a wide spectrum, containing frequency components spanning from 100 to 300 T, undergoes characteristic changes with decreasing thickness. Physically, the SdH frequency $B_F$ is directly linked to the area enclosed by the electron orbits on the Fermi surface $\mathcal{A}_k(E_F)$ described by the semiclassical relation, $B_F = \frac{\hbar}{2\pi e}\mathcal{A}_k(E_F)$, where $\hbar$ is the reduced Plank's constant, and $e$ is the elementary charge. The presence of multiple frequencies is therefore attributed to distinct electronic pockets $\alpha_{1,2,\cdots,i}$. The SdH oscillation patterns, stretching to higher frequencies with the increase of thickness, can be categorized into three main groups. Beginning with the simplest single distinct component $\alpha_1$ in samples thinner than 12 nm (Fig. 1f), it progresses to the intermediate thickness range of 26 and 19 nm, exhibiting two components $\alpha_1$ and $\alpha_2$, and further extends to three or more oscillation frequencies in thick samples (32 and 60 nm).

Despite its layered crystal stacking, bulk Bi$_2$O$_2$Se still has a typical 3D Fermi surface with significant dispersion along $\Gamma - Z$ direction of the Brillouin zone (BZ)[15]. Using the generic relation, $m_z^* = \hbar^2 \left(\frac{d^2 E}{d k_z^2}\right)^{-1}$, at the bottom of the conduction band, the effective mass out of B-O plane ($k_z$ along $\Gamma - Z$) yields $m_z^* = 0.34\, m_0$ [20,24]. The substantial $k_z$ dispersion sets the Bi$_2$O$_2$Se apart from conventional VdW semiconductors, where the $k_z$ dispersion is much weaker. As typical $m_z^*$ values can well exceed $m_0$ in conventional VdW materials, a pronounced layer degree of freedom with intrinsic quasi-2D electronic states can already appear in the bulk. In contrast, a 2D state in Bi$_2$O$_2$Se has to be built by quantizing the dispersion along $k_z$ direction.

At 4.2 K, we estimate the thermal de Broglie wavelength $\lambda_D = \sqrt{\frac{2\pi\hbar^2}{m_\perp^* k_B T}}$ of the conduction electrons to be 62.4 nm. Since the electrons in Bi$_2$O$_2$Se are highly delocalized, the extended wavefunction in the out-of-plane direction renders the dispersion of electrons susceptible to the sample thickness. In the case of electron motion along the *z*-direction restricted by the crystal boundaries, the energy spectrum becomes quasi-continuous, $E = E_i + \frac{\hbar^2 k_\parallel^2}{2 m_\parallel^*}$, where $k_\parallel$ is an in-plane wavevector. The $E_i$ quantizes as $E_{1,2...i}$, where the energy separation $\Delta E_{i,i+1} = E_{i+1} - E_i$ is determined by the specific shape of the confining potential [1,3]. Using the simplest well-type potential, the energy spacing reduces to $\Delta E_{i,i+1} = \left(\frac{\hbar^2\pi^2}{2m_\perp^*}\right)\frac{2i+1}{d^2}$, where $d$ is the quantum-well width (Fig. 1g). In $Bi_2O_2Se$ flakes thinner than the characteristic length $\lambda_D$, quantization leads to the formation of concentrically arranged 2D sub-bands. We label the energy at the bottom of each sub-band as $E_i$, as illustrated in the simplified sub-band structure in Fig. 1g.

**Quantum Hall effect in a two-dimensional electron gas**

Based on the quantization scheme discussed above, the critical prerequisite for realizing the QHE is to form a 2DEG with sufficiently large sub-band separation. Therefore, we choose the sample thickness as the only device parameter and focus on thin samples below ~15 nm. In these thicknesses, where the estimated 2D sub-band separation can well exceed the thermal broadening ($\Delta E_{i,i+1} \gg k_B T$ at 4.2 K), we can then use the gate voltage $V_G$ to adjust the Fermi level near the bottom of conduction band and selectively occupy the lowest sub-band ($i = 1$) of the 2DEG[1].

Applying a strong magnetic field perpendicular to this 2DEG can further quantize the energy spectrum as

$$E = E_1 + \left(l + \frac{1}{2}\right)\frac{\hbar eB}{m_\parallel^*} + m_s g^* \boldsymbol{\mu}_B \cdot \boldsymbol{B}, \qquad (1)$$

where $l = 0, 1, 2, \cdots$ is the Landau level (LL) index, $m_s = \pm\frac{1}{2}$ is the spin quantum number, $g^*$ the Landé factor (see also Supplementary Note 1 and 6), and $\mu_B$ the Bohr magneton. Fig. 2a shows a set of MRs measured at 1.34 K up to 35 T for various electron concentrations in an 11 nm sample. When $B$ is higher than ~25 T and carrier concentrations $n_{2D}$ is tuned $< 3.8\times10^{12}$ cm$^{-2}$, the SdH oscillation develop into a QHE with a small residual $\rho_{xx}$ and a well-quantized plateaus in $\rho_{xy}$ at even integer fractions of $R_K = \frac{h}{ve^2}$, where $v = 2, 4, 6\cdots$, are the filling factors corresponding to the lowest filled LL of the bottommost $E_1$ sub-band (Fig. 2b). Contributions from the next sub-band, $E_2$, becomes increasingly noticeable in the SdH oscillations at $n_{2D} > 5.02\times10^{12}$ cm$^{-2}$, where multicomponent oscillations initiate at $B > 15$ T (see Supplementary Note 3).

We proceed with further characterization of our device by extracting crucial parameters that enable QHE. Fig. 2c shows the $n_{2D}$ and mobility $\mu_H$, extracted from Hall effect measurements (see Supplementary Note 2). Here, the electron concentration follows a linear dependence with gate voltage over a wide range of $V_G$ as $\Delta n_{2D} = \frac{\varepsilon_0 \varepsilon_r}{ed_{ox}}\Delta V_G$, where $\varepsilon_0$ is the permittivity of a vacuum. The $SiO_2$ gate capacitor used in our device has a thickness $d_{ox} = 300$ nm and a dielectric constant $\varepsilon_r = 3.87$. For the thin $Bi_2O_2Se$ device $\mu_H$ peaks around ~2200 cm$^2$V$^{-1}$s$^{-1}$ at $n_{2D} \approx 3.5\times10^{12}$ cm$^{-2}$ and saturates at higher concentrations. The $\mu_H$ then drops rapidly toward a percolation limit below the critical density $n_c$[25] (see Supplementary Note 2).

The amplitude of the primary SdH oscillations $\alpha_1$ exhibits a pronounced temperature dependence (Extended Data Fig.3), which can be described by the Lifshitz-Kosevich equation $R_T = \frac{\lambda T}{sinh(\lambda T)}$. By using $\lambda = \frac{2\pi^2 k_B}{e\hbar B}m_c$, we can extract the cyclotron mass, $m_c$. Two distinct electronic states with $n_{2D}$ = 6.94 and 1.62×10$^{12}$ cm$^{-2}$ display different rates of amplitude decay as a function of $T$ (Extended Data Fig.4), yielding $m_c \approx 0.27$ and $0.13m_0$, respectively. Cyclotron masses $m_c$ extracted at different carrier concentrations $n_{2D}$ (Fig. 2e), demonstrate an overall tendency to decrease towards the band edge. Notably, the smallest $m_c = 0.13m_0$ obtained from the SdH is in good agreement with the reported values of $m^*$, measured at the band edge of Bi$_2$O$_2$Se crystals[20,23,24].

**Sub-band splitting in a confined electron system**

We further investigate the 2D sub-bands in all other flakes thinner than 15 nm, by analyzing the dependence of the primary SdH oscillation frequency $B_F$ as a function of $n_{2D}$. The relationship between $B_F$ and $V_G$ for different thicknesses is established from standard fan diagrams. The frequency of oscillation in resistivity is a function of the density of carriers $n_{SdH}$ as $B_F = \frac{\phi_0 n_{SdH}}{g_s \Delta_i}$, where $\phi_0$ is the magnetic flux quantum, and $g_s \Delta_i$ are spin and sub-band degeneracies. As shown in Fig. 3a, $n_{2D}$ and $B_F/\phi_0$ follow a linear relationship, with $g_s \Delta_i = 2$ up to a characteristic carrier density $n_0$, determined individually for each given thickness. The $g_s \Delta_i = 2$ factor originates from the single spin-degenerate sub-band $\Delta_1$ situated at the $\Gamma$ point of the first BZ[15]. The factor also aligns well with the even integer fractions observed from the Hall plateaus (Fig. 2b). With the increase in $n_{2D}$, an abrupt change of linear dependence between $n_{2D}$ and $B_F/\phi_0$ is universally observed when $n_{2D} > n_0$, for each given thickness. For example, in the 11 nm sample, the slope is nearly doubled at $n_{2D} > n_0 \sim 3.5 \times 10^{12}$ cm$^{-2}$. This abrupt change is caused by a doubling of the sub-band index, $\Delta_i = 1 \rightarrow 2$, in the density of states $D(E) = g_s \Delta_i \frac{m_\parallel^*}{2\pi\hbar^2}$ when the Fermi level enters the higher sub-band $E_2$.

Considering the first two sub-bands of a quantum well with an infinite potential wall, the energy separation is given by $\Delta E_{1,2} = \frac{3\hbar^2 \pi^2}{2m_\perp^*}\frac{1}{d^2}$. To raise $E_F$ from $E_1$ to $E_2$, we can calculate the required additional carriers as $n_e = D(E)(E_2 - E_1) = \frac{3}{2}\left(\frac{m_\parallel^*}{m_\perp^*}\right)\frac{1}{d^2}$, which is proportional to a mass anisotropy ratio $\sim \frac{m_\parallel^*}{m_\perp^*}$, and inversely proportional to sample thickness $\sim \frac{1}{d^2}$. We find that the thickness dependence of the onset carrier density $n_0$ can be well described by the quantum well model (Fig. 3b), with $m_\parallel^*/m_\perp^* = 0.24/0.3$. It is worth noting that, although the free-electron analysis (Fig. 3c) does not consider the sophisticated shape of the quantum well caused by the interplay between sample thickness and electric confinement potential, this straightforward model can already be used as a guide to select a proper $d$ value as the sole design parameter for CVD-based $Bi_2O_2Se$ when applying it to QHE devices.

**Landau level structure at high magnetic field**

We now turn to the high field regime, where the Landau levels can be well resolved. By simultaneous measuring $\rho_{xx}$ and $\rho_{xy}$ as a function of $V_G$ (Fig. 4a) at a field of 30 T, pronounced QHE plateaus are observed at $\rho_{xy} = \frac{h}{ve^2}$, where the integer filling factor $v$ undergoes successive changes by multiples of 2 each time the Fermi level crosses a Landau Level (LL). The plateaus in $\rho_{xy}$ are accompanied by equidistantly spaced minima in $\rho_{xx}$ with a separation of $\Delta V_G$ ~21 V, this spacing further supports the notion that $g_s \Delta_i = 2$. Considering additional $n_{2D}$ accumulated by $\Delta V_G$ (Fig. 4a), we can calculate the filling required to occupy individual LL as $(\alpha \Delta V_G) = g_s \Delta_i B/\phi_0$, when we choose $\alpha = 7.13 \times 10^{12}$ cm$^{-2}$ and $g_s \Delta_i = 2$.

The sequential filling of Landau levels significantly changes at $V_G \geq 70$ V, where $n_{2D}$ starts to exceed $n_0$ for the 11 nm thick sample. The inset of Fig. 4a shows this change in the filling pattern in $\sigma_{xy}$ at different magnetic fields. For the sequence of $\sigma_{xy} = N\left(\frac{2e^2}{h}\right)$, where $N$ is the total number of LLs crossed, the filling alters at 15 and 22.5 T when the sub-band degeneracy incriminator increases from $\Delta_i = 1 \to 2$. Namely, the interval doubles at $N = 5$ from 1, 2, 3, 4, 5 to 7, 9, 11 (See also Extended Data Fig.5 and 6). This doubling of LL degeneracy disappears at larger cyclotron energy $\hbar\omega_c$ (30 T), where the filling reverts to the 1, 2, 3, 4, 5 sequence. The change in the filling pattern can be explained by the reconstructed sub-band alignment between $E_1$ and $E_2$ as shown in Fig. 4b or from the reconstructed Landau fan diagram (see Supplementary Note 5). Here, at $B = 15$ T, the energy separation is $\Delta E_{1,2} \approx 4.8\ \hbar\omega_c$, therefore, the zeroth LL of the second sub-band $l_2 = 0$ is closely aligned in energy with the fifth LL of the first sub-band $l_1 = 5$. Whereas, at $B = 30$ T, the $\Delta E_{1,2}$ becomes $\approx 2.4\ \hbar\omega_c$, hence, the $l_2 = 0$ LL aligns in between $l_1 = 2$ and 3 (Extended Data Fig.7). The additional carrier required to reach $E_2$ can be estimated as $n_e = 2.4 \frac{eB}{\pi\hbar} = 3.48 \times 10^{12}$ cm$^{-2}$ (Supplementary Note 5), which is closely agrees with the $n_0$ obtained from Fig. 3b.

**Conclusions**

In summary, we have successfully demonstrated the QHE in the one or two-subband 2DEG of a CVD–grown oxide – $Bi_2O_2Se$. Despite its inherent 3D electronic structure, we found that samples thinner than the $\lambda_D$ can function as field-effect transistors. The gated $Bi_2O_2Se$ conforms to a simple quantum well model based on the free electron approximation, with the layer thickness serving as the sole device design parameter. Due to its inherently small effective mass compared to typical high-mobility oxides, the QHE effect in $Bi_2O_2Se$ can be observed even in samples with very low mobilities, offering significant flexibility for utilizing the quantum Hall effect in oxide materials. The flexible layered structure of this new oxide family and its unique small mass makes it a promising platform for exploring the versatile quantum Hall effect in oxide materials.

**Methods**

**Crystal growth**

Single crystals of $Bi_2O_2Se$ were grown using conventional chemical vapor deposition (CVD) on *f*-mica substrates. Two powder-based precursors, $Bi_2O_3$ and $Bi_2Se_3$, were vaporized at high temperatures in a multizone tube furnace and carried by an Ar flow in a quartz tube downstream to the mica substrates, which were held at lower temperatures. The growth process was conducted in an under-pressure environment of ~400 Torr. The high temperatures used to evaporate sources were raised to 620 °C at a rate of 25 K/min. The isotherm for CVD growth was held for 40 minutes. The growth is terminated by a natural cooling[12].

**Device fabrication**

Single crystals of $Bi_2O_2Se$ with uniform morphology ranging from 6 to 60 nm in thickness were carefully chosen for device fabrication. Square-shaped flakes, having 20-40 μm in lateral size, were selected and transferred using a polybisphenol carbonate (PC) adhesive layer, which can be dissolved in chloroform within 20 minutes. The AFM topography of transferred samples on $SiO_2/Si^{++}$ substrates is shown in Extended Data Fig.1. All AFM scans were conducted right after the PC-assisted transfer from the mica. We used conventional e-beam lithography to fabricate the Hall-bar devices. As shown in the inset of each AFM micrograph, the contacts to the devices consist of Pt/Au electrodes, with a thickness of 5/35 nm.

**Transport measurements**

The low-temperature electrical transport was measured in a He-4 cryostat. We use standard lock-in amplifiers (SR-860) to probe the transport in a four-probe configuration. The carrier concentration of the Hall-bar device is tuned by gating through a 300 nm dielectric layer of thermally grown $SiO_2$. A Keithley 2611B source meter was used to apply the DC gate bias.

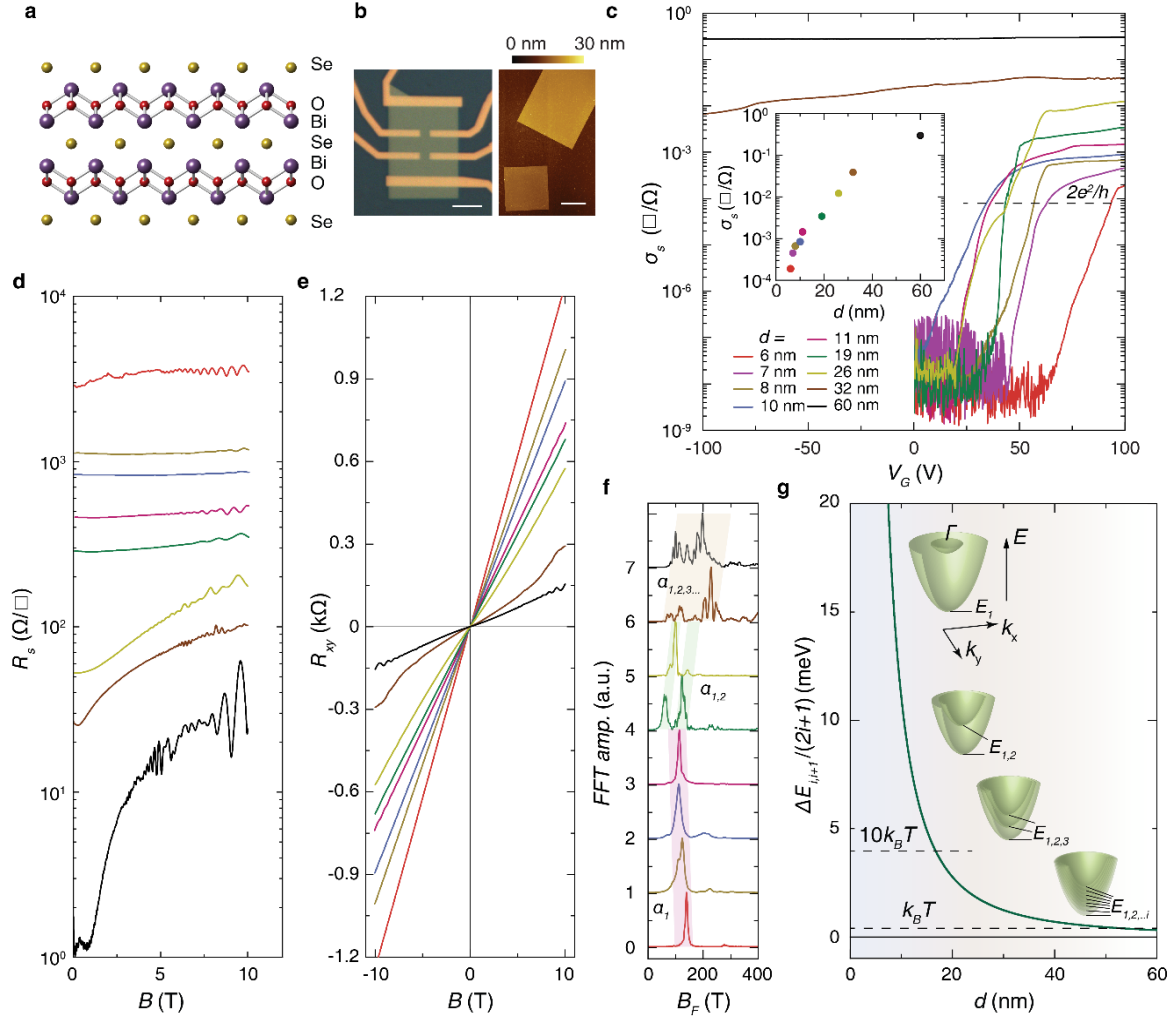

**Fig. 1| Crystal structure and electronic properties of Bi$_2$O$_2$Se thin films.**

**a,** Side view of the Bi$_2$O$_2$Se bilayer. The single crystal is bonded via a chalcogen atomic plane interlaced between layers of covalently bonded bismuth-oxide. **b,** Optical micrograph of a 10 nm thick Bi$_2$O$_2$Se device and its atomic force microscopy (AFM) image before fabrication. The scalebar is 5 µm. **c,** Sheet conductivity $\sigma_s$ as a function of gate voltage $V_G$ at $T$ = 4.2 K. Samples with different thicknesses $d$ are labeled with different colors. We use the same color scheme throughout the whole paper to trace the different thicknesses. The dashed line marks the quantum conductance $\frac{2e^2}{h}$. *Inset* shows the $\sigma_s$ for the saturation conductivity, namely the "ON" states, as a function of device thicknesses. **d** and **e** Shubnikov-de Haas oscillations (SdH) of Bi$_2$O$_2$Se flakes in the "ON" state **(c)** and the corresponding Hall effect. The data is plotted by taking the symmetric $\frac{1}{2}[R_{xx}(+B) + R_{xx}(-B)]$ and antisymmetric $\frac{1}{2}[R_{xy}(+B) - R_{xy}(-B)]$ components of longitudinal and transverse resistance, respectively. **f,** Fast Fourier transform (FFT) spectra of the SdH oscillation in the inverse magnetic field domain 1/$B$. **g,** Schematic structure of sub-bands configuration of the electronic pocket located at $\Gamma$ point of the first Brillouin zone of Bi$_2$O$_2$Se starting from a thick sample to its thin counterpart (*top to bottom*). The bottom of each sub-band is indicated as $E_{1,2,\cdots,i}$, where $i$ is a sub-band index. Dashed lines mark the thermal broadening energy at 4.2 K.

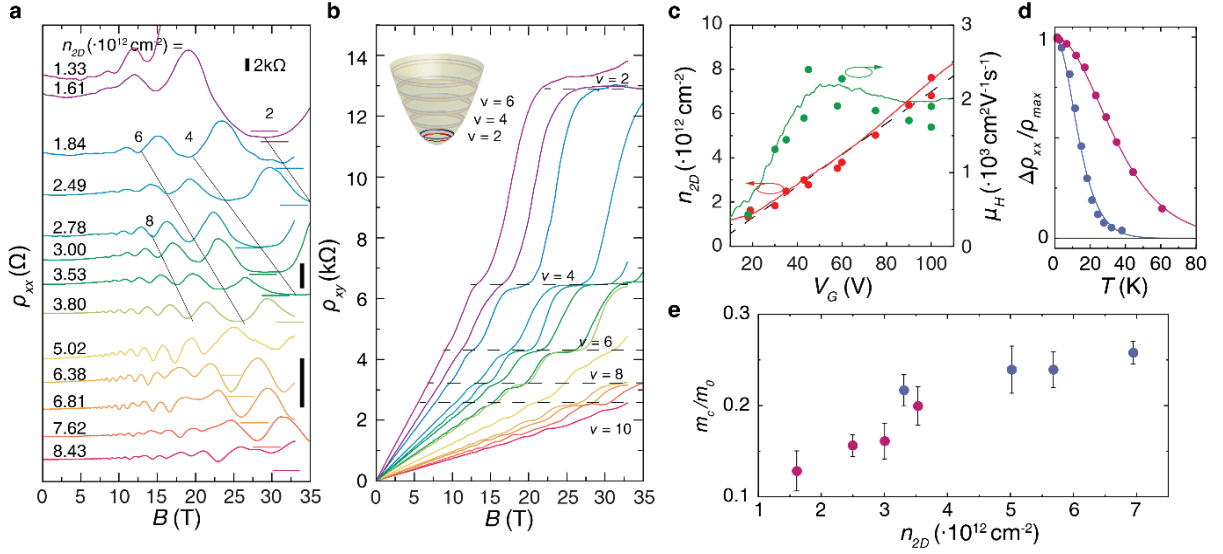

**Fig. 2| Shubnikov-de Haas oscillations and quantum Hall effect.**

**a,** Symmetric component of longitudinal resistivity $\rho_{xx}$ of a 11 nm Bi$_2$O$_2$Se flake at carrier concentrations $n_{2D}$ ranging from 1.33 to 8.43×10$^{12}$ cm$^{-2}$. The magnetic field dependences of $\rho_{xx}$ are offset for clarity. The short horizontal bars, marking $\rho_{xx} = 0$ for each curve, are color-coded accordingly, and the height of all scale bars is 2 kΩ. From shortest to longest, three scale bars correspond to carrier density ranges up to 1.61, 3.80, and 8.43×10$^{12}$ cm$^{-2}$, respectively. The resistance minima are labeled as $\nu$ corresponding to different filling factors. **b,** Hall effect $\rho_{xy}$ measured at $T = 1.36$ K simultaneously with $\rho_{xx}$ shown in panel **(a)**. Dashed lines mark the position of even fractions of the von Klitzing constant $R_K = \frac{h}{\nu e^2}$, where $\nu = 2, 4, 6, 8 \cdots$. Plateaus observed at even integer filling factors correspond to filled spin-split Landau Levels (LLs) of the $\Gamma$ – band (*see inset*). **c,** Concentration of induced electrons $n_{2D}$ (*left axis*) and Hall mobility $\mu_H$ (*right axis*) as a function of gate voltage $V_G$. The solid line and the filled circles correspond to the two data sets, obtained from varying the carrier concentration at magnetic field $B = 2$ T and vice versa. The dashed line is a fitting to a capacitor model of SiO$_2$ gate dielectric (300 nm). **d,** The decay of the amplitude of the SdH oscillation as a function of temperature. We plot two exemplary states measured at 30 Tesla, having $n_{2D} = 1.61$ and 6.94×10$^{12}$ cm$^{-2}$. The Lifshitz-Kosevich fittings to the data (solid lines) yield $m_c \approx 0.13$ and 0.27 $m_0$, respectively. **e,** Cyclotron mass $m_c/m_0$ extracted at different electron concentrations for two different samples 11 nm (*pink dot*) and 10 nm thick (*blue dot*).

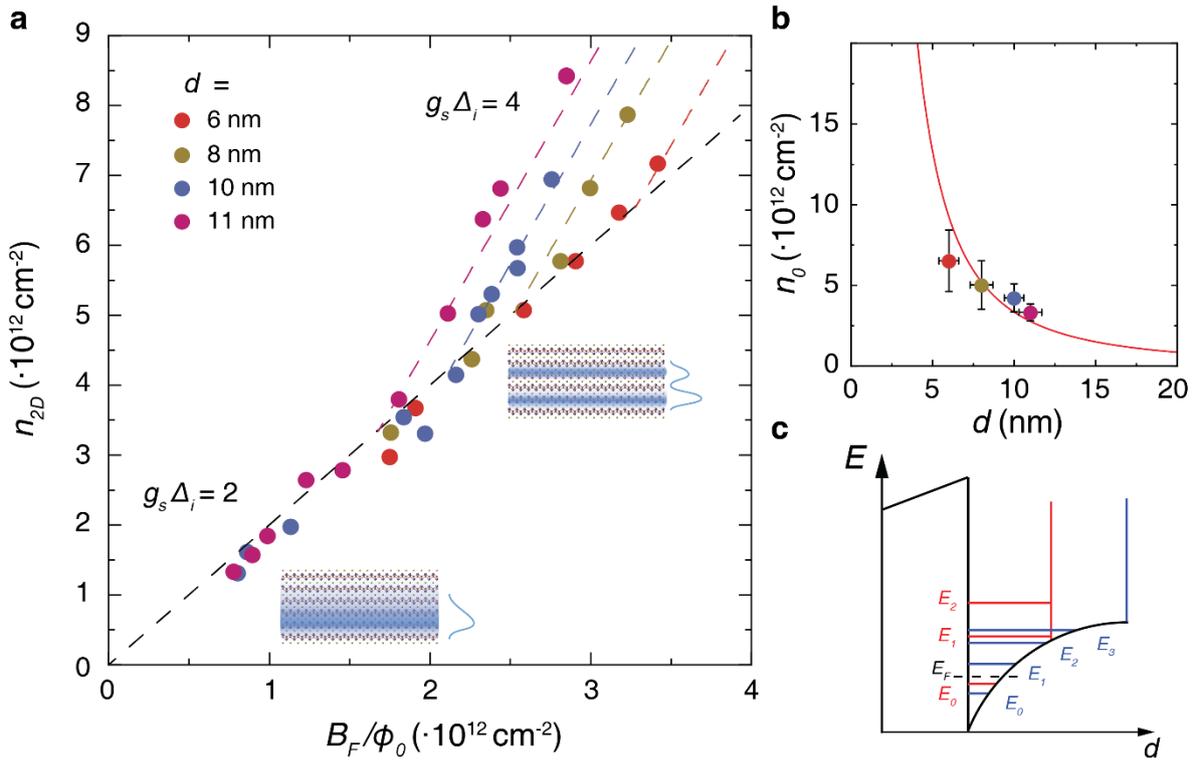

**Fig. 3| Quantum confinement in thin films.**

**a,** Hall effect carrier concentration $n_{2D}$ is plotted as a function of $B_F/\phi_0$, where $B_F$ is extracted from the SdH oscillation. The $n_{2D}$ values are color-coded for different thicknesses. The black dashed line is a fitting with the LL degeneracy factor $g_s\Delta_i = 2$. The colored dashed lines guide the deflections from the degeneracy factor $g_s\Delta_i = 2$ to $4$ at different critical Hall carrier densities $n_0$, for different thicknesses. **b,** Critical carrier concentration $n_0$, marking the onset of occupying the second sub-band as a function of sample thickness. **c,** Confinement potential tuned by the gate voltage for two different thicknesses. The thinner sample acquires greater sub-band splitting energy between $E_i$ and $E_{i+1}$.

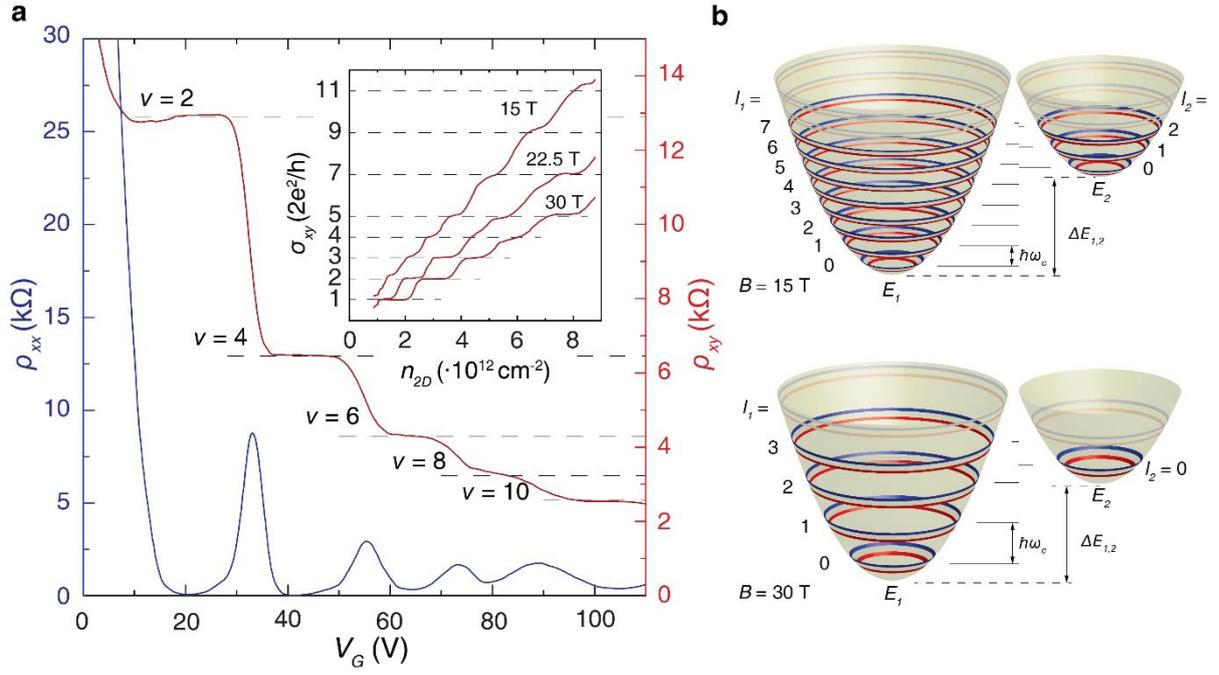

**Fig. 4| Quantum Hall effect in 2DEG.**

**a,** Longitudinal $\rho_{xx}$ and Hall resistivity $\rho_{xy}$ as a function of applied gate voltage at 30 T and T = 1.32 K. The sample is 11 nm thick. At 30 T, the LLs are periodically spaced in $\Delta V_G$ up to ~70 V. Considering that the oscillations sequentially appear at $\Delta V_G \approx 21\,V$ ($B = 30\,\text{T}$) and the relation between LL occupation and LL energy separation is $(\alpha \cdot \Delta V_G) = g_s \Delta_i \cdot B/\phi_0$, we obtain $g_s = 2$. Above ~70 V, the oscillation period rearranges due to populating a higher sub-band $\Delta_i = 1 \rightarrow 2$. *The inset* shows transverse conductivity $\sigma_{xy}$, in the units of $\frac{2e^2}{h}$, plotted versus the measured Hall carrier density at 15, 22.5, and 30 T, respectively. **b,** Schematic picture of the sub-band alignment in a magnetic field of 15 (*top*) and 30 T (*bottom*) at the highest accessed occupation. The sub-bands are shifted horizontally for clarity. The LL index is denoted as $l$, unfilled levels are semitransparent.